\documentclass[preprint,aps,pra,showpacs,floats,superscriptaddress]{revtex4}

\usepackage{amsmath}
\usepackage{amsfonts}
\usepackage{amssymb}
\usepackage[german,english]{babel}
\usepackage{bm,pstricks}                          % bold math

\newcommand{\lmin}{\lambda_{\text{min}}}
\newcommand{\rhob}{\overline{\rho}}
\newcommand{\rhot}{\widetilde{\rho}}
\newcommand{\tmin}{\tau_{\rm min}}
\newcommand{\tH}{\tau_{\rm H}}

\newcommand{\ue}{\text{e}}

\def\eea{\end{eqnarray}}
\def\bea{\begin{eqnarray}}
\def\ee{\end{equation}}
\def\be{\begin{equation}}

\begin{document}

\title{Constraint on periodic orbits of chaotic systems given by Random Matrix Theory}

\author{Alejandro G. Monastra}

\address{Gerencia Investigaci\'on y Aplicaciones, Comisi\'on Nacional de Energ\'\i a At\'omica, Avda. General Paz 1499, (1650) San Mart\'\i n, Argentina}
\address{Consejo Nacional de Investigaciones Cient\'\i ficas y T\'ecnicas, Avda. Rivadavia 1917, (1033) Buenos Aires, Argentina}

\date{\today}

\begin{abstract}
Considering the fluctuations of spectral functions, we prove that if chaotic systems fulfill the Bohigas-Gianonni-Schmit (BGS) conjecture, which relates their spectral statistics to that of random matrices, therefore by virtue of Gutzwiller trace formula, the instability of classical periodic orbits is constrained. In particular for two-dimensional chaotic systems, the Lyapunov exponent $\lambda_p$ of each periodic orbit $p$ should be bigger than a minimum value $\lambda_{\text{min}} \geq 0.850738$. This opens the possibility of new constraints for a system to be fully chaotic, or the failure of the BGS conjecture.
\end{abstract}

\pacs{05.45.Mt, 03.65.Sq}

\maketitle

%%%%%%%%%%%%%%%%%%%%%%%%%%%%%%%%%%%%%%%%%%%%%%%%%%%%%%%%%%%%%%%%%%%%%%%%%
\section{Introduction}
%%%%%%%%%%%%%%%%%%%%%%%%%%%%%%%%%%%%%%%%%%%%%%%%%%%%%%%%%%%%%%%%%%%%%%%%%

A difficult task in classical mechanics is to prove that a dynamical system is fully chaotic. Except for some few maps \cite{cat, baker} and billiards \cite{sinai, bunimovich, BalVor} where analytical proofs exist, the numerical exploration of phase-space is the only practical way to decide this question. However, small regions of regular motion around stable periodic points can escape the numerical analysis, invalidating the method.

One way to approach this question is to study the periodic orbits of the system. Linearization of equations of motion near a periodic orbit $p$ is characterized by the monodromy matrix $M_p$. In an autonomous Hamiltonian system with $d$ degrees of freedom, the dimension of the monodromy matrix is $2d-2$, and the modulus of the determinant is one. Eigenvalues of this matrix are usually written in the form $e^{\lambda_p^{(i)}}$, where $\lambda_p^{(i)}$ are the so called Lyapunov exponents, for $1 \leq i \leq 2d-2$. If the modulus of at least one of the eigenvalues is bigger than one (or equivalently the real part of some $\lambda_p^{(i)}$ is positive), then the periodic orbit is unstable and the dynamics in the vicinity is hyperbolic. In a fully chaotic system all periodic orbits should be unstable, and the dynamics is fully hyperbolic \cite{descon}.

In particular, for two dimensional Hamiltonian systems, the eigenvalues of the monodromy matrix of a periodic orbit $p$ take the simple form $\{e^{\lambda_p},\pm e^{-\lambda_p} \}$. Hence full hyperbolicity implies that for every periodic orbit $p$, $\lambda_p$ should have a nonzero real part.

On the other hand, quantization of Hamiltonian systems have shown that imprints of classical dynamics are present in both  eigenfunctions \cite{BerryRW} and eigenvalues. These relationships are the main subject of study of Quantum Chaos. Regarding the eigenvalues, two important conjectures have been well established:

(i) Berry-Tabor (BT) conjecture \cite{BerryTabor}: for {\it generic} regular systems the eigenvalues of its quantization are completely uncorrelated and their statistics is Poissonian.

(ii) Bohigas-Giannoni-Schmit (BGS) conjecture \cite{BGS}: in fully chaotic systems the eigenvalues are correlated (in particular there exists level repulsion) and they show the same statistical properties as those of random matrix ensembles. Particularly, for systems with time reversal symmetry, the fluctuations are those of the Gaussian Orthogonal Ensemble (GOE). For systems without time reversal symmetry the comparison is done with the Gaussian Unitary Ensemble (GUE)

The BT conjecture was corroborated numerically for several cases and analytically for some of them. Nevertheless some counterexamples were found, and the problem remains in the definition of a {\it generic} regular system. The BGS conjecture was also checked numerically in many theoretical systems, specially quantum billiards \cite{QBilliards1, QBilliards2} and even experimentally \cite{Hatom, Richter}. We stress that the connection between complex dynamics and Random Matrix Theory (RMT) started in the 1950's, where it was applied to the statistics of resonances in atomic nuclei \cite{nuclei}. The conjecture was also proved in some specific (but less realistic) systems like graphs \cite{quantumgraph} and billiards on negative curvature surfaces \cite{BalVor}. Although for the latter case some counterexamples were found depending on the tiling properties of the billiards \cite{BGGS}.

In the way to prove these conjectures is essential to use the connection between classical dynamics and eigenvalues that exist using the semiclassical approximation. For sufficiently high energies the density of states

\begin{equation}
\rho (E) = \sum_j \delta (E-E_j) \ ,
\label{DofS}
\end{equation}
splits into a smooth plus an oscillatory part $\rho = \rhob + \rhot$. The smooth part admits an expansion in powers of energy called the Weyl series, which depends on global properties of the system such as volume, surface and topology of the classical phase space, not regarding the integrable or chaotic dynamics. The oscillatory part writes as

\begin{equation}
\rhot (E) = 2 \sum_p A_p (E) \cos \left[ S_p (E)/\hbar+ \nu_p \right] \ ,
\label{Gutz}
\end{equation}
where the sum is made over all classical periodic orbits $p$ of the system that exist at energy $E$, $S_p$ is the corresponding action of the periodic orbit of period $\tau_p = {\rm d} S_p / {\rm d} E $, and $\nu_p$ is a phase called the Maslov index which depends on the number of turns of the stable and unstable varieties along the periodic orbit.

The value of the amplitude $A_p$ depends on the stability of the periodic orbit. Berry and Tabor \cite{BerryTabor} found a general formula for stable periodic orbits in terms of the actions and frequencies of the classical tori that support the orbit. If the orbit is unstable, the amplitude is given by the Gutzwiller formula \cite{Gutzwiller}

\begin{equation}
A_p = \frac{\tau_p}{2\pi\hbar \ r_p \sqrt{\left| \det (M_p - I) \right|}} \ ,
\label{Apchaotic}
\end{equation}
in terms of the period $\tau_p$, the monodromy matrix $M_p$ ($I$ is the identity matrix), and the repetition  number $r_p$ when the orbit $p$ is non primitive ($r_p=1$ for primitive periodic orbits).

The semiclassical expansion (\ref{Gutz}) can at least explain qualitatively some aspects of the above mentioned conjectures. In general, around a stable periodic orbit there exists a continuous family of periodic orbits, all with similar periods and actions. The contribution of a family produces a large fluctuation on the density of states if we compare it with the contribution of an unstable periodic orbit, which is always isolated. Therefore, in integrable systems, large fluctuations of the energy levels are expected, allowing for the characteristic clustering of a Poissonian distribution predicted by BT conjecture. On the other hand, in fully chaotic systems, where all the periodic orbits are unstable, the spectrum has less fluctuations and more rigidity, a characteristic shared with random-matrix eigenvalues.

We show in this work an even deeper relation between the BGS conjecture and the classical dynamics. For every 2-D fully chaotic system:

$\bullet$ if there is at least one periodic orbit whose Lyapunov exponent $\lambda_p$ verifies $0 < \lambda_p < \lmin$, then the fluctuations of the quantum spectrum do not obey random matrix statistics;

$\bullet$ if the quantum spectrum has random matrix fluctuations, then all classical periodic orbits have Lyapunov exponents $\lambda_p > \lmin$.

The two propositions can be demonstrated in the same way, but they have different implications. The former implies that the BGS conjecture is false, and more conditions are needed for a chaotic system to have a spectrum with random-matrix fluctuations. The latter has implications on classical dynamics: not only all Lyapunov exponents should be positive, moreover they should have a minimum value.

In order to prove these propositions, a general spectral function depending on two parameters is defined in Section \ref{SpectralFunctions} and its fluctuations are estimated semiclassically by periodic orbit theory. In section \ref{Variance} the variance of fluctuations is calculated in terms of the spectral form factor, deducing the constraint to the monodromy matrix for each periodic orbit. We analize the implications of this result, particularly on the Lyapunov exponents of periodic orbits of two-dimensional systems, giving some conclusions in Section \ref{Conclusions}.

%%%%%%%%%%%%%%%%%%%%%%%%%%%%%%%%%%%%%%%%%%%%%%%%%%%%%%%%%%%%%%%%%%%%%%%%
\section{Spectral functions} \label{SpectralFunctions}
%%%%%%%%%%%%%%%%%%%%%%%%%%%%%%%%%%%%%%%%%%%%%%%%%%%%%%%%%%%%%%%%%%%%%%%%

Let $\{ 0 \leq E_1 \leq E_2 \leq \cdots E_j \cdots \}$ be a discrete spectrum coming from the quantization of a bounded Hamiltonian system. We define a spectral function depending on two real positive parameters with energy units

\begin{equation}
F(\mu, T) = \sum_{j=1}^{\infty} f \left( \frac{\mu - E_j}{T} \right) \ ,
\label{Fdef}
\end{equation}
where $f(x)$ is a continuous square integrable function. We assume for simplicity that $f(x)$ is an even function, with a Fourier transform

\begin{equation}
\hat{f} (\omega) = \int_{-\infty}^{\infty} f(x) \cos (\omega x) ~ {\rm d}x \ .
\end{equation}
We ask for $f(x)$ to be a localized function, as a Gaussian, that decays exponentially for $|x| \gg 1$. If $f(x) = \log (1+ e^{x})/(1+e^{x})+\log (1+ e^{-x})/(1+e^{-x})$, the spectral function $F$ corresponds to the grand canonical entropy of a non-interacting Fermi gas with chemical potential $\mu$, temperature $T$ (measured in units of Boltzmann constant), confined in a potential with a single-particle spectrum $\{ E_j \}$ \cite{Landau}.

By definition, the sum (\ref{Fdef}) can be written as the energy integral of $f$ times the density of states (\ref{DofS})

\begin{equation}
F(\mu, T) = \int_{-\infty}^{\infty} f \left( \frac{\mu - E}{T} \right) \rho (E) ~ {\rm d}E = T \int_{-\infty}^{\infty} f (x) ~ \rho (\mu - T x)  ~ {\rm d}x \ ,
\label{Fintegral}
\end{equation}
where the integration variable $E$ has been changed to $x = (\mu - E)/T$. Because $f(x)$ is localized around $x=0$, it is only necessary to know the density of states in a window of order $T$ around the energy $\mu$. Moreover, if $\mu$ is high in the spectrum and $T \ll \mu$, we can introduce the semiclassical expansion (\ref{Gutz}) for $\rho$. With these conditions, in the energy window where the integrand (\ref{Fintegral}) is significant, all classical quantities needed in (\ref{Gutz}) remain almost constant. The main dependence on energy comes from the phase of the cosine, where the action $S_p$ is divided by $\hbar$, arriving to

\begin{equation}
\rho (\mu - T x) \approx \rhob (\mu) + 2 \sum_p A_p (\mu) \cos \left[ \frac{S_p (\mu) - \tau_p (\mu) T x}{ \hbar} + \nu_p \right] \ .
\end{equation}
Introducing this approximation into (\ref{Fintegral}), the integral can be performed analytically and the spectral function $F$ also splits into a smooth plus an oscillatory part

\begin{eqnarray}
\overline{F} (\mu,T) &=& I_0 \ T \ \rhob(\mu) \ ,  \label{Fsmooth} \\
\widetilde{F} (\mu,T) &=& 2 \ T \sum_p A_p ~ \hat{f} \left( \frac{T \tau_p}{\hbar} \right) \cos \left( \frac{S_p}{\hbar} + \nu_p\right) \ ,
\label{Fosc}
\end{eqnarray}
where $I_0 = \int_{-\infty}^{\infty} f (x)  {\rm d}x = \hat{f} (0) $ is a dimensionless number. In the latter expression, all classical quantities $A_p, S_p, \tau_p$ and $\nu_p$ are computed at energy $E=\mu$.  Formula (\ref{Fosc}) is the main result of this section. Its structure is similar to semiclassical formula (\ref{Gutz}), with $\mu$ playing the role of the energy. However, if $\hat{f} (\omega)$ decays exponentially for $\omega \gg 1$, the prefactor $T \hat{f} ( T \tau_p /\hbar )$ introduces a cut-off for long periodic orbits. Therefore, series (\ref{Fosc}) for $\widetilde{F}$ can be convergent if parameter $T$ is big enough to compensate the exponential proliferation of periodic orbits in a chaotic system (which is related to its topological entropy).

%%%%%%%%%%%%%%%%%%%%%%%%%%%%%%%%%%%%%%%%%%%%%%%%%%%%%%%%%%%%%%%%%%%%%%%%
\section{Variance of the fluctuations} \label{Variance}
%%%%%%%%%%%%%%%%%%%%%%%%%%%%%%%%%%%%%%%%%%%%%%%%%%%%%%%%%%%%%%%%%%%%%%%%

For a fixed value of $T$, rapid fluctuations of $\widetilde{F}$ are observed as $\mu$ is varied. These fluctuations have some distribution and typical size depending on several parameters. To characterize the fluctuations around some value $\mu$ it is necessary to define some averaging procedure. For any spectral function $G$ that depends on the energy $\mu$, the average is defined as

\begin{equation}
\left\langle G (\mu) \right\rangle = \frac{1}{\Delta \mu} \int_{\mu-\Delta \mu/2}^{\mu+\Delta \mu/2} G (\mu'){\rm d}\mu' \ ,
\end{equation}
where $\Delta \mu$ is the size of the window where the fluctuations are analyzed. From Eq. (\ref{Fosc}), every periodic orbit defines an energy scale $\delta \mu_p = h / \tau_p$ where the phase of the cosine varies $2 \pi$. The largest of these scales is $E_c = h / \tmin$, coming from the shortest periodic orbit of period $\tmin$. Then, if $\Delta \mu \gg E_c$ we know that $\langle \cos (S_p(\mu) / \hbar) \rangle = 0$ for every periodic orbit. On the other hand, if $\Delta \mu \ll \mu$ the classical quantities are almost constant inside the energy window (see \cite{LebMon2002} for a detailed discussion on the involved energy scales). From now on we consider the fluctuations only at sufficiently high energies $\mu$ where the two conditions are fulfilled

\begin{equation}
E_c \ll \Delta \mu \ll \mu \ ,
\end{equation}
implying $\langle \widetilde{F} \rangle = 0$. Using Eq.~(\ref{Fosc}), the variance $\langle \widetilde{F}^2 \rangle$ is written as the average of a double sum over periodic orbits. The product of cosines may be expressed as one half the sum of the cosine of the sum of the actions and the sum of their difference. The term with the sum of actions vanishes by the average, arriving to

\begin{equation}
\left\langle \widetilde{F}^2 \right\rangle = 2 T^2 \left\langle \sum_{p,p'} A_p A_{p'} \hat{f} (\tau_p T/\hbar) \hat{f} (\tau_{p'} T/\hbar) \cos \left( \frac{S_p - S_{p'}}{\hbar} \right) \right\rangle \ .
\end{equation}
We can relate the variance with the semiclassical expression of the form factor (the Fourier transform of the two point correlation function of the density of states) \cite{Berry1985}

\begin{equation}
K(E,\tau) = h^2 \left\langle \sum_{p,p'} A_p A_{p'} \cos \left( \frac{S_p - S_{p'}}{\hbar} \right) \delta \left( \tau - \frac{\tau_p + \tau_{p'}}{2} \right)  \right\rangle \ .
\label{Ksem}
\end{equation}
Given this expression, the variance of the fluctuations is written in the integral form
\begin{equation}
\left\langle \widetilde{F}^2 \right\rangle = \frac{T^2}{2 \pi^2 \hbar^2} \int_{0}^{\infty} \hat{f}^2 (\tau T/\hbar) K(\mu, \tau) {\rm d}\tau \ .
\end{equation}

All the dynamical information of the system is now contained in the form factor. It is also known that for times $\tau \ll \tH$, where $\tH = h \overline{\rho} (E)$ is the Heisenberg time, the only contributions to the form factor come from terms where $p'=p$, and $p'=p^T$ if the orbit $p$ has a time reversal partner $p^T$. This is the so called diagonal approximation, and the form factor simplifies to

\begin{equation}
K_D (E,\tau) \approx  h^2 \sum_{p} g_p \ A_p^2 \ \delta( \tau - \tau_p ) \ .
\label{Kdiag}
\end{equation}
In systems with time-reversal symmetry $g_p=2$ for all periodic orbits (excepting those few ones that are self-retracing, where $g_p=1$). If time-reversal symmetry is broken, $g_p=1$ for all periodic orbits. For chaotic systems, by Hannay-Ozorio de Almeida sum rule \cite{HanOzo}, this simplifies to $K_D(\mu,\tau) = g \ \tau$ for $\tau \gg \tmin$ ($g=2$ for time-reversal systems, $g=1$ if the symmetry is broken). This approximation is valid for the full form factor up to a time much smaller than Heisenberg time. Thereafter, off-diagonal corrections become important and expression (\ref{Ksem}) converge to the random matrix form factor

\begin{eqnarray}
K_{\mbox{\tiny GOE}}(E,\tau) &=& \left\{
\begin{array}{ll}
2 \tau - \tau \log \left( 2 \frac{\tau }{ \tH} +1 \right)   \ , & \tau < \tH \\
& \\
2 \tH  - \tau \log \left( \frac{2 \tau + \tH }{ 2 \tau - \tH } \right)  \ , & \tau > \tH
\end{array}
\right. \\
&& \nonumber \\
K_{\mbox{\tiny GUE}}(E,\tau) &=& \left\{
\begin{array}{ll}
\tau \ , & \tau < \tH \\
& \\
\tH \ , & \tau > \tH
\end{array}
\right.
\end{eqnarray}
GOE (GUE) should be used for chaotic systems with (without) time reversal symmetry. For $\tau \gg \tH$ the form factor goes to a constant value $\tH$ for both symmetries.

If we consider the contribution to the variance of $\widetilde{F}$ given by all long periodic orbits $\tau_p \geq \tau_1$, such that $\tmin \ll \tau_1 \ll \tH$, we have

\begin{equation}
V (\tau_1) \equiv \frac{T^2}{2 \pi^2 \hbar^2} \int_{\tau_1}^{\infty} \hat{f}^2 (\tau T/\hbar) K(\mu, \tau) {\rm d}\tau \ .
\end{equation}
For this time regime the form factor should correspond to the random matrix prediction. Moreover, we know that $K_{\mbox{\tiny RMT}}(\mu, \tau) < K_D(\mu,\tau) = g \tau$, arriving to the inequality

\begin{equation}
V (\tau_1) < \frac{g T^2}{2 \pi^2 \hbar^2} \int_{\tau_1}^{\infty} \tau \hat{f}^2 (\tau T/\hbar) {\rm d}\tau \ .
\end{equation}
Changing the integrating variable to $\omega = \tau T / \hbar$ the integral simplifies to

\begin{equation}
V (\tau_1) < \frac{g}{2 \pi^2} \int_{\omega_1}^{\infty} \omega \hat{f}^2 (\omega) {\rm d}\omega \ ,
\end{equation}
where $\omega_1 = \tau_1 T / \hbar$. Because the integrand is always finite and positive, we can extend the lower limit up to zero

\begin{equation}
V (\tau_1) < \frac{g}{2 \pi^2} \int_{0}^{\infty} \omega \hat{f}^2 (\omega) {\rm d}\omega = \frac{g}{2 \pi^2} \hat{I}_1 \ .
\label{Vmaximum}
\end{equation}
The integral converges to $ \hat{I}_1$, assuming that $\hat{f} (\omega)$ decays rapidly enough, which imposes further restrictions on the possible primitive functions $f(x)$ that can be used. The contribution to the variance of long periodic orbits has an upper limit that depends only on the Fourier transform of the function $f(x)$.

On the other hand, using Gutzwiller formula (\ref{Apchaotic}) for $A_p$ in chaotic systems, we can write the contribution to the variance from long periodic orbits in the semiclassical form
\begin{equation}
V (\tau_1) = \frac{1}{2 \pi^2} \sum_{\tau_p > \tau_1} \frac{g_p \ \omega_p^2 \ \hat{f}^2 (\omega_p) }{r_p^2 \left| \det (M_p - I) \right|} \ ,
\label{Vsum}
\end{equation}
where $\omega_p = \tau_p T / \hbar$. In this sum, all terms  are positive. For some $\omega = \omega^*$ the function $\omega^2 \ \hat{f}^2 (\omega)$ has a finite maximum value $I^*$. Therefore, if we consider the contribution of one periodic orbit $p$  to the sum (\ref{Vsum}), and we fix the parameter $T$ to $T^* = \omega^* \hbar / \tau_p$, we show that

\begin{equation}
V (\tau_1) > \frac{g_p \ I^* }{ 2 \pi^2 r_p^2 \left| \det (M_p - I) \right|} \ . \label{Vminimum}
\end{equation}

Comparing this to the inequality (\ref{Vmaximum}), that holds for every $T$, we arrive to the condition

\begin{equation}
\frac{g_p }{ 2 \pi^2} \frac{\ I^* }{ r_p^2 \left| \det (M_p - I) \right|} < \frac{g}{2 \pi^2} \hat{I}_1 \ .
\label{Inequality1}
\end{equation}
For the general case of primitive periodic orbits ($r_p=1$), and non self-retracing orbits ($g_p=g$), condition (\ref{Inequality1}) simplifies to

\begin{equation}
\left| \det (M_p - I) \right| > \alpha = \frac{I^*}{\hat{I}_1} \ .
\label{Inequality2}
\end{equation}
This is the key result of the work, showing a constraint on the instability of periodic orbits in chaotic systems. For two-dimensional systems the Lyapunov exponents of an unstable periodic orbit are $\{ \lambda_p, -\lambda_p \} $ by the condition that $ | \det M_p | = 1$. Therefore

\begin{equation}
\left| \det (M_p - I) \right| = 4 \sinh^2 ( \lambda_p / 2) \ ,
\label{determinant}
\end{equation}
and inequality (\ref{Inequality2}) can be inverted giving

\begin{equation}
\lambda_p > 2 \ \text{arcsinh} ( \sqrt{ \alpha} / 2 ) \ .
\label{Inequality3}
\end{equation}
This condition should be fulfill for every even symmetric function $f(x)$ with finite $I_0$ integral and finite $\hat I_1$ integral related to its Fourier transform. We face a difficult variational problem: to find the function that maximizes the value of $\alpha$, in order to find the threshold value $\lambda_{\text{min}}$ for the Lyapunov exponent in 2D systems. A first test with some elementary functions has given a maximum value $\alpha = 2/ \ue = 0.73576$ for the Gaussian function $f(x) = \ue^{-x^2}$. From this result we explored the family function $f(x) = \ue^{-|x|^n}$, varying the exponent $n$. By a numerical analysis we found the maximum value $\alpha = 0.76847$ for exponent $n=3$, and using (\ref{Inequality3}) the highest limit to the Lyapunov exponent is at least

\begin{equation}
\lambda_p > \lambda_{\text{min}} \geq 0.850738 \ldots \ .
\label{lyapunov_condition}
\end{equation}
Some other families of functions were tested although none of them had given a higher value of $\alpha$.

%=======================================================================
\section{Conclusion} \label{Conclusions}
%=======================================================================

We have shown that if BGS conjecture is true, i.e. spectral fluctuations of fully chaotic systems follow RMT, then instability of classical periodic orbits has a constraint. Particularly in 2D systems Lyapunov exponents of long periodic orbits have a minimum value.

As it was mentioned in Section \ref{SpectralFunctions}, a physical example of a spectral function $F(\mu, T)$ is the grand canonical entropy of a non-interacting fermion gas in a confining potential. For this case, a detailed analysis of the behavior of $\left\langle \widetilde{F}^2 \right\rangle$ as a function of parameter $T$ (the temperature) was done in \cite{LebMon2002} for gases with chaotic single-particle motion. For $T \ll \Delta$ the variance of entropy fluctuations grows linearly with $T$. Increasing $T$ up to few mean level spacings, the variance saturates to the maximum value $V_{\text{max}} = g  \hat{I}_1 /(2 \pi^2)$ given by RMT, see Eq.~(\ref{Vmaximum}). When $T$ is of order $E_c$, the shortest periodic orbits become relevant and an exponential decreasing of the variance is expected.

This analysis will be qualitatively similar for other localized functions $f(x)$. Therefore, even if the proof of (\ref{Inequality2}) was made for long periodic orbits $ \tau_p \gg \tmin$, where usually the Lyapunov exponents are large, we conjecture that this condition can be extended to the shortest periodic orbits. Violation of (\ref{Inequality2}) for a short periodic orbit $p$ would imply that the variance of the spectral function $F(\mu, T)$, after reaching the maximum value $V_{\text{max}}$, instead of decreasing exponentially to zero, would continue to grow beyond the value given by Eq.~(\ref{Vminimum}) at some value $T^*= \omega^* \hbar / \tau_p$. Numerical calculations in \cite{LebMon2002} for a Sinai-type billiard, and in \cite{Riemannium} for a fictitious system given by the non-trivial zeros of the Riemann Zeta function, have shown that the variance of entropy fluctuations, as a function of temperature, never become bigger than the limit imposed by RMT. We induce from these observations that monodromy matrix of all periodic orbits in these systems fulfill condition (\ref{Inequality2}).

Nevertheless, an extensive numerical analysis of other chaotic systems is needed to give more evidence to the conjecture for short periodic orbits. Or eventually a counterexample would keep the constrain $\left| \det (M_p - I) \right| > 0.76847$ only for long periodic orbits, as it was proved in this work.

\section*{Acknowledgments}
The author wants to thank Marcos Saraceno for useful discussions, and CONICET and CNEA for financial support.

%\section*{References}

\end{document}